\documentclass[aps, prl,amsmath,amssymb, reprint, superscriptaddress]{revtex4-2}

\usepackage{graphicx}% Include figure files
\usepackage{dcolumn}% Align table columns on decimal point
\usepackage{bm}% bold math
\usepackage[ruled]{algorithm2e}
\usepackage{color}

\begin{document}

\preprint{APS/123-QED}

\title{A numerical model for time-multiplexed Ising machines\\based on delay-line oscillators.}

\author{Roman V. Ovcharov}
\affiliation{ 
Department of Physics, University of Gothenburg, Gothenburg 41296, Sweden
}

\author{Victor H. González}
\affiliation{ 
Department of Physics, University of Gothenburg, Gothenburg 41296, Sweden
}

\author{Artem Litvinenko}%
\affiliation{ 
Department of Physics, University of Gothenburg, Gothenburg 41296, Sweden
}

\author{Johan \AA kerman}
\affiliation{ 
Department of Physics, University of Gothenburg, Gothenburg 41296, Sweden
}
\affiliation{
Center for Science and Innovation in Spintronics and Research Institute of Electrical Communication, Tohoku University, Aoba-ku, Sendai 980-8577, Japan
}

\author{Roman S.  Khymyn}%
\affiliation{ 
Department of Physics, University of Gothenburg, Gothenburg 41296, Sweden
}

\date{\today}

\begin{abstract}
Ising machines (IM) have recently been proposed as unconventional hardware-based computation accelerators for solving NP-hard problems. In this work, we present a model for a time-multiplexed IM based on the nonlinear oscillations in a delay line-based resonator and numerically study the effects that the circuit parameters, specifically the compression gain $\beta_r$ and frequency nonlinearity $\beta_i$, have on the IM solutions. We find that the likelihood of reaching the global minimum---the global minimum probability (GMP)---is the highest for a certain range of $\beta_r$ and $\beta_i$ located near the edge of the synchronization region of the oscillators. The optimal range remains unchanged for all tested coupling topologies and network connections. We also observe a sharp transition line in the ($\beta_i, \beta_r$) space above which the GMP falls to zero. In all cases, small variations in the natural frequency of the oscillators do not modify the results, allowing us to extend this model to realistic systems.
\end{abstract}

\maketitle

Ising machines (IM) are a type of cost function minimization embedding that allows one to obtain optimal or near-optimal solutions to NP-hard problems such as graph partitioning~\cite{Ushijima2017-graph-partitioning-d-wave}, circuit layout design~\cite{Barahone1988-circuit-design}, graph colouring~\cite{OIM-Wang2019}, condition satisfiability~\cite{Sharma2023-augmented-IM}, traveling salesman~\cite{Sutton2017-optimization-nanomagnets}, and weighted knapsack~\cite{Ibarra1975-knapsack}. Recently, several experimental realizations of IMs were demonstrated~\cite{Mohseni2022-IM-review, Gonzalez2024-spintronic-devices-next-gen}, which can be divided into two types: spatially resolved, where each graph edge and node has a designated place in space, and time-multiplexed, where each node, i.e., Ising spin, is processed sequentially in time. The latter approach is gaining traction thanks to its excellent scalability to large numbers of spins and connections. In this work, we expand upon our universal numerical model for such time-multiplexed IMs, originally formulated to describe spinwave-based IMs~\cite{litvinenko2022SWIM}, by introducing a larger number of oscillators and non-trivial couplings. We study the model's response to variations in both amplitude and frequency nonlinearities to find the best conditions for convergence and high solution accuracy. Our findings can be applied to a large range of time-multiplexed IM designs and can help realize the potential of these emerging hardware-based solvers.

\begin{figure}[ht!]
\centering
	\includegraphics[width=0.98\linewidth]{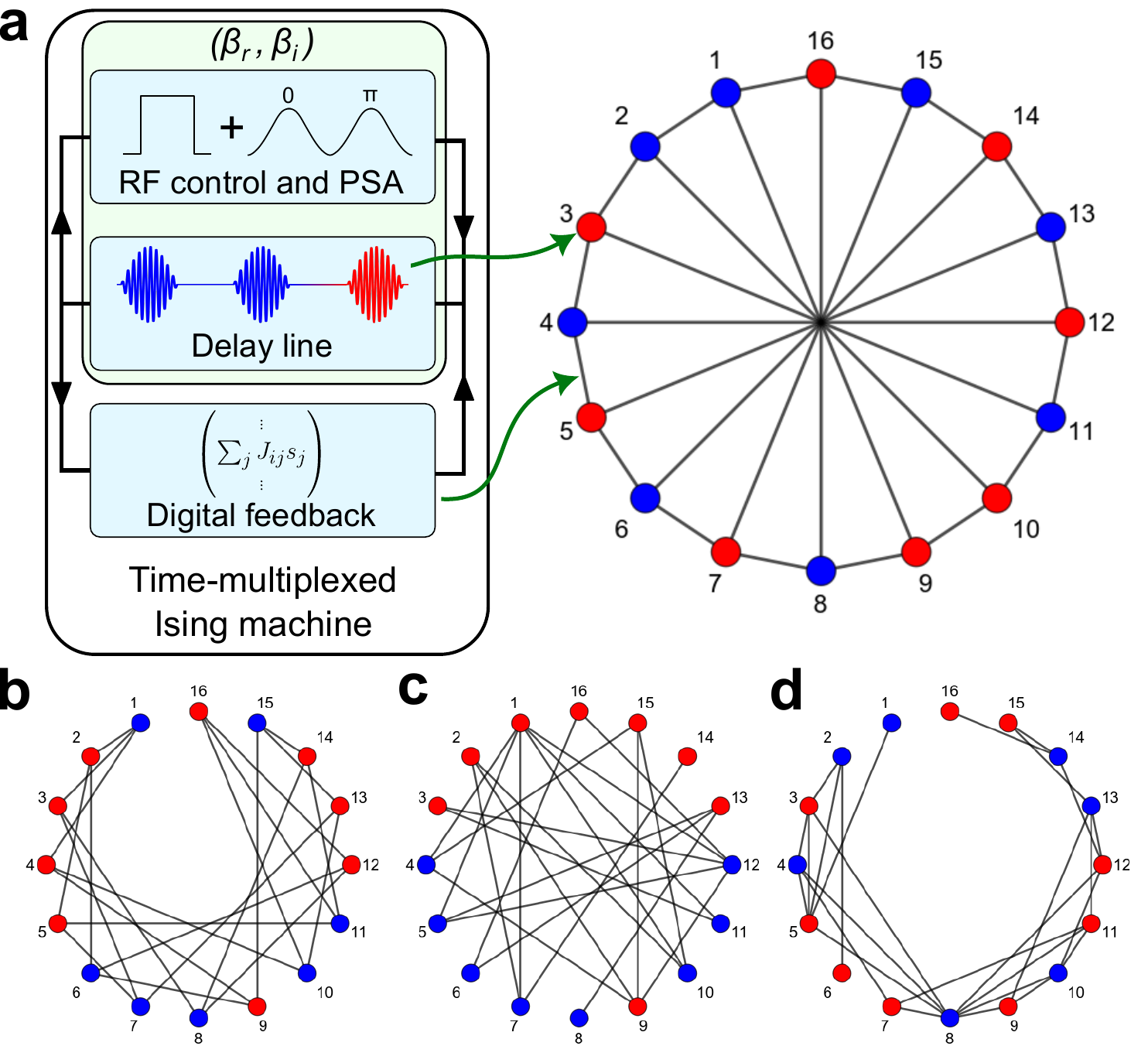}
	\caption{\textbf{Schematic of a time-multiplexed IM and the graphs chosen to study the influence of compression gain $\beta_r$ and frequency nonlinearity $\beta_i$ on the delayed IM model.} (a) The green arrows point to the electronic element associated with the respective term in the Ising Hamiltonian. The spins are constructed using RF pulses in a delay line and coupled using digital feedback. The graph here is the M{\"o}bius ladder. (b-d) Other graphs studied with the numerical model.}
	\label{fig:matrices}
\end{figure}

Much effort has been dedicated to the development of IMs using a variety of hardware embeddings and models~\cite{Syed2023-physics-bifurcation-optimizers}: CMOS logic cells~\cite{tsukamoto2017accelerator, yamaoka2015-20k-spins}, simulated annealing algorithms~\cite{Earl-2005-parallel-tempering, Kanao2022-simulated-bifurcation}, magnetic~\cite{Sutton2017-optimization-nanomagnets, albertsson2021ultrafastSTNOIM, Houshang2022-SHNO-IM, McGoldrick2022-SHNO-IM} and electrical~\cite{Afoakwa-BRIM} oscillators, magnetic tunnel~\cite{Sutton2017-optimization-nanomagnets} and superconducting junctions~\cite{albash2018-d-wave-computer, boixo2016-multiqubit-tunnelling-quantum-annealers}, spin~\cite{litvinenko2022SWIM}, light~ \cite{McMahon2016-100-CIM, Honjo2021-100k-CIM, Wang2013-CIM-DOPOs, Utsunomiya2011-CIM, Hamerly2019-CIM-vs-QA, Stroev2023-analog-photonics-computing}, light-matter~\cite{Kalinin2020-polaritonic-xy-machine} solitons, and even dedicated atoms~\cite{Kiraly2021-atomic-boltzmann-machine}. Interest in this field comes from IMs' efficiency in finding optimal or near-optimal solutions for problems with configuration spaces that grow exponentially with problem size. Such problems are called nondeterministic polynomial (NP) because their solutions cannot be found by simple combinatorics in polynomial time and are of considerable interest to a wide variety of fields, including economics, logistics, biochemistry, meteorology, and condensed matter physics. At their core, Ising machines are solvers that find the lowest energy state of the Ising Hamiltonian:
\begin{equation}
    \label{eq:ising-hamiltonian}
    H = -\sum_{i\neq j} J_{ij} s_i s_j + \sum_{i} h_i s_i
\end{equation}
where the $J_{ij}$ terms are the couplings between pairs of spins of a set $s=(s_1, s_2, ..., s_n)$ with $s_i = \{-1, +1\}$, and $h_i$ are the individual biases of each one. It is well known that finding the ground state of Eq.~(\ref{eq:ising-hamiltonian}) is an NP-hard problem, and there exist mathematical transformations that encode a majority of NP problems into an Ising one. 

For an IM with a cost function of the form of Eq.~(\ref{eq:ising-hamiltonian}), one must ensure the binarization of its spins, a high density of their pairwise couplings (there can be up to $n(n-1)/2$ spin-spin connections) and individual bias for each one  (up to $n$ biases). Achieving a high coupling density is a challenge for spatially resolved machines, since hardware constraints limit the coupling density. Any general Ising problem must typically be sparsified by adding auxiliary spins to reduce the average coupling density and fit the hardware's topology~\cite{Aadit2022SparseIM}.

Time multiplexing, a well-known technique in signal processing, is an alternative to sparsification that can embed the original problem into hardware with the aid of an external coupling element. In this arrangement, the artificial spins are constructed using amplitude modulation of light or RF signals and phase-sensitive amplification to produce phase-binarized pulses propagating in a loop where they do not interact, as shown in the green box in Fig.~\ref{fig:matrices}(a). In the RF case, this loop is composed of the electronic elements necessary to produce the artificial spins and a delay line, where the spins are hosted and propagated. The nonlinear properties of the loop are essential for the IM operation as it has been shown that control over the nonlinearity in IMs can improve their operation speed and solution quality~\cite{Inui2022-amplitude-control-CIM, Gonzalez2023-Global-SWIM, Bohm2021}. Using our model, we systematically study the role of two nonlinear parameters---the gain compression $\beta_r$ and the nonlinear frequency shift $\beta_i$---in the dynamics of an IM and how they should be chosen for optimal operation.

At a certain point in the loop, the propagation path is split, and part of the signal is input into the coupling element, where the \textit{local field} is calculated. The local field of each spin is the weighted sum of all other spin contributions plus its individual bias, which is then re-injected into the loop using digital feedback and added to the original signal exactly after one circulation around the ring. The mixing of the signals is the coupling mechanism, and it results in the minimization of a cost function that can be set with arbitrarily coupled spins, as shown in Fig.~\ref{fig:matrices}(a). The time-multiplexing technique is very versatile as has been demonstrated with different kinds of pulses, including light solitons~\cite{Utsunomiya2011-CIM, Hamerly2019-CIM-vs-QA, Honjo2021-100k-CIM}, spin waves~\cite{litvinenko2022SWIM,Gonzalez2023-Global-SWIM} and surface acoustic waves~\cite{litvinenko2023SAWIM}. 

\section{Model}

To study the dynamics of the IM, we used the oscillator-based Ising machine (OIM) formalism to construct a network of oscillators that can minimize Eq.~(\ref{eq:ising-hamiltonian})~\cite{OIM-Wang2019}. In OIM, the artificial spin state is constructed using a network of coupled oscillators, the phase of which is binarized by second harmonic injection locking (SHIL) \cite{carpentieri2013SHIL, albertsson2021ultrafastSTNOIM}. It has been shown that SHIL can be used for mapping the Ising Hamiltonian into an oscillator network while its phase dynamics minimize an encoded cost function \cite{OIM-Wang2019}, with the phases $\{0, \pi\}$ representing the $\{1, -1\}$ spins values.

The evolution of each RF pulse, which represents an Ising spin, can be considered as a time-multiplexed separate, weakly nonlinear oscillator with a delay in a feedback loop. The complex amplitude $c_j$ of the $j-$th Ising spin can be described by the left part of the equation derived in \cite{tiberkevich2014sensitivity}:
\begin{multline}
\frac{dc_j}{dt}+(i \omega_0+\Gamma_0)\ c_j-K\left[1-\beta_r\frac{p_j(t-\tau)-p_0}{p_0}\right] \\
\times \exp\left[-i\beta_i\frac{p_j(t-\tau)-p_0}{p_0}\right]c_{j}(t-\tau)=\xi_j(t),
\label{eq:delay}
\end{multline}
where $\tau=\tau_{delay}$ is a delay time along the loop, i.e., in the delay line, $\Gamma_0$ describes losses, $\omega_0$ is the resonant frequency of the loop, $K$ -- amplification gain and $\beta_r$ -- coefficient of the gain compression, $\beta_i$ defines the nonlinear frequency shift, and $p(t)=|c(t)|^2$ is a power of oscillations with an operating point at $p_0$.  

The right part of Eq.(\ref{eq:delay}) contains additional signals from the outer loops, namely the synchronization signal at the double frequency and couplings from other spins:
\begin{equation}
    \xi_j(t)=K_e\exp^{i\omega_e t}c_j^*+\kappa \sum_{i\neq j} J_{ij}\frac{c_{i}}{|c_{i}|},
    \label{eq:coupling-sum}
\end{equation}
where $\omega_e \simeq 2\omega_0$ represents the binarizing SHIL signal and $\kappa$ is a coupling coefficient to other spins.

To determine the roles that $\beta_r$ and $\beta_i$ play in the steady state solution space of a general time-multiplexed IM, we used coupling schemes with different symmetries, energy minima, and optimal solutions, as shown in Fig.~\ref{fig:matrices}. The M{\"o}bius ladder (Fig.~\ref{fig:matrices}(a) and the graph shown in Fig.~\ref{fig:matrices}(b) have been previously used to benchmark the performance of time-multiplexed IM~\cite{McMahon2016-100-CIM}, while the other two graphs (c,d) are randomly generated.

\begin{figure}[hbt!]
    \centering
    \includegraphics[width=\linewidth]{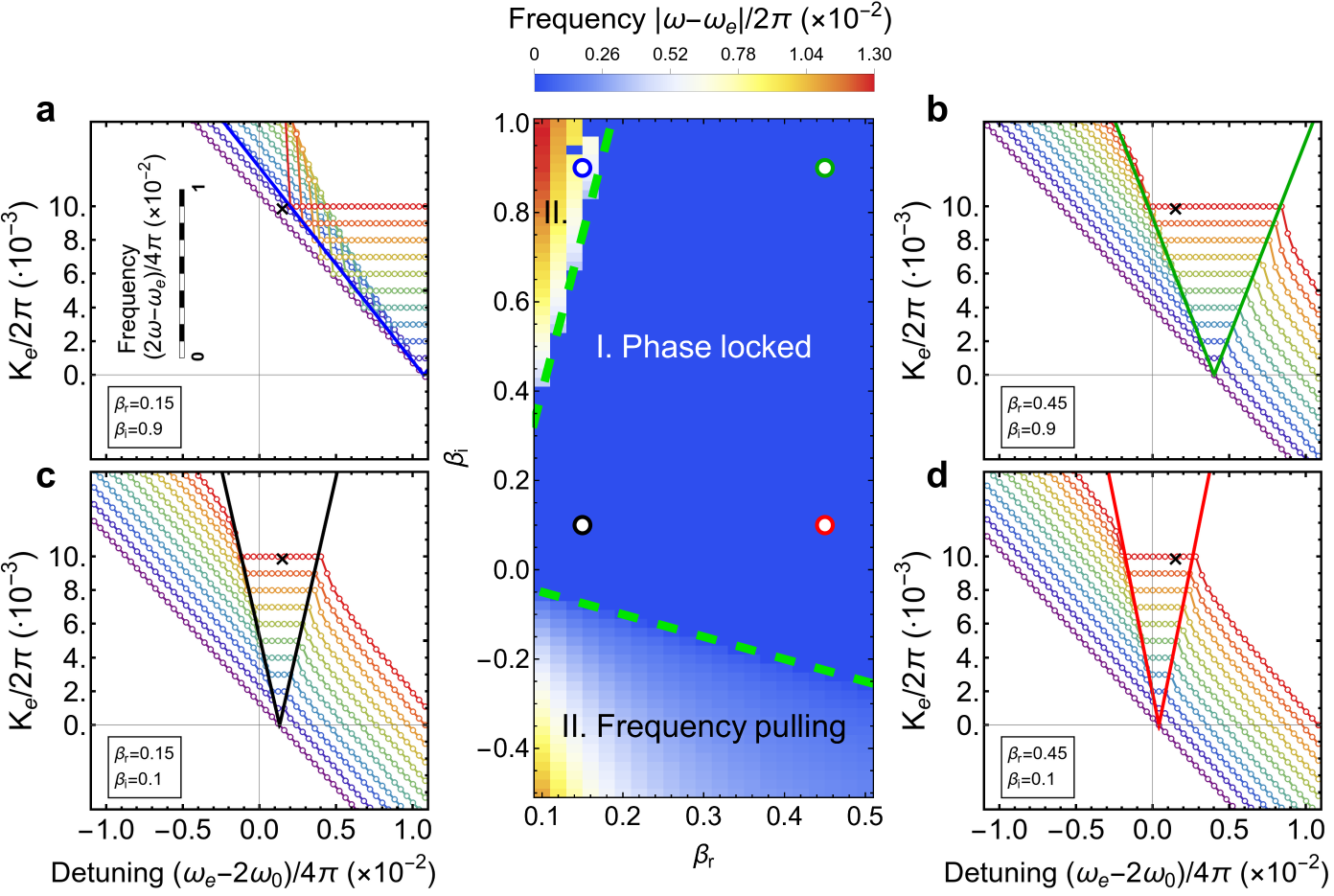}
    \caption{\textbf{Role of $K_e$.} The central figure shows the synchronization map in the $(\beta_r,\beta_i)$ space for the chosen $K_e=0.01$. The dashed lines denote the edges between synchronized (I) and unsynchronized (II) regions. Each $(\beta_r,\beta_i)$ tuple changes the locking bandwidth of the oscillators. (a-d) Frequency of the oscillator $2\omega-\omega_e$ vs SHIL detuning $\omega_e-2\omega_0$ and strength $K_e$ for four points in our region of interest; the color of the Arnold tongues corresponds to the color of the marker in the locking map.}
    \label{fig:coupling-search}
\end{figure}

Plugging Eq.~(\ref{eq:coupling-sum}) into Eq.~(\ref{eq:delay}), we obtain a system of $n$ coupled ordinary differential equations describing the complex amplitude of the spins. We numerically solved them using Wolfram Mathematica with the following default parameters: $\omega_0/2 \pi = 1.0$, $\omega_e/2 \pi = 2\times1.0015$, $\tau = 10$, $\Gamma_0/2\pi = 0.05$, $K / 2 \pi = 0.06$, $\kappa / 2 \pi = 0.003$. Besides the amplitude and phase nonlinearities, the coupling strength $K_e$ to the binarizing signal plays an important role in the injection locking of the oscillators. To keep the results consistent, we performed a sweep of $K_e$ and found a value for which the oscillators lock to the binarizing signal for most ($\beta_r$, $\beta_i$) tuples in the $\beta$-plane.

\section{Results and discussion}

\begin{figure*}[hbt!]
\centering
	\includegraphics[width=0.99\linewidth]{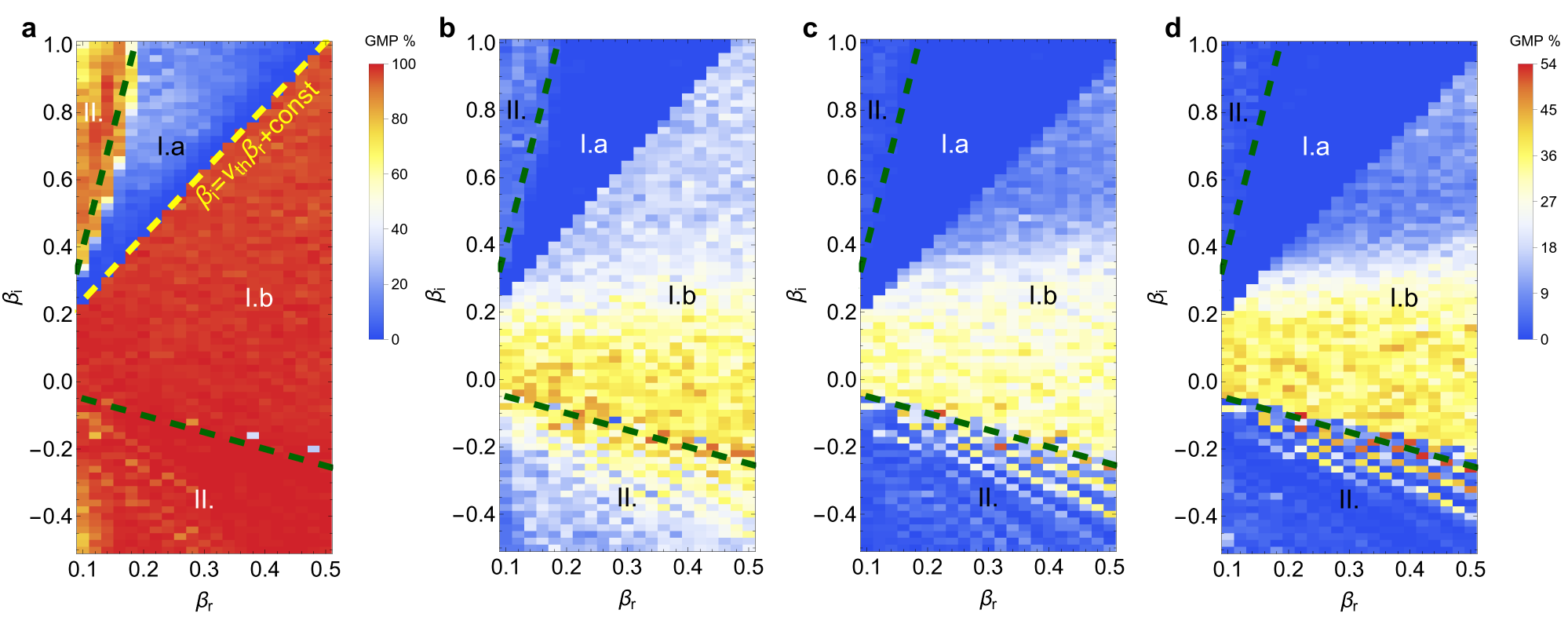}
	\caption{\textbf{Global minimum probabilities (GMP) for the 4 graphs.} Each sub-figure \textbf{(a-d)} corresponds to the graph in Fig.~\ref{fig:matrices} with the same letter. In all cases, we observe a sharp transition where the GMP drops to zero above a certain threshold and non-zero GMP in the upper left region, despite a lack of synchronization to $\omega_e$. For non-trivial cases (b-d), we observe that an uptick in GMP appears near the edge of the synchronization region, denoted by the lower dotted line.}
	\label{fig:result}
\end{figure*}

First, we calibrated the binarizing strength $K_e$ for the robust phase-locking of the oscillators in the wide range of ($\beta_r$, $\beta_i$). The response dynamics of the oscillators \emph{w.r.t.}~the binarizing signal are shown in the insets (a-d) of Fig.~\ref{fig:coupling-search} for the four points located in the $\beta$-plane. We observe that the locking range of the oscillators (i.e., the region where $\omega = \omega_e/2$) increases linearly with $K_e$, as shown by the Arnold tongues \cite{carpentieri2013SHIL, litvinenko2021analog, Rajabali2023Injection} (the triangle-shaped regions) in (a-d). The binarizing coupling strength $K_e/2\pi=0.01$ was chosen as it allowed us to explore both synchronized and unsynchronized regimes on the $\beta$-plane for all coupling matrices. In the colormap, we show the frequency locking of an oscillator for all points in the $\beta$-plane between $\beta_r \in [0.1,0.5]$ and $\beta_i \in [0,1]$ with a step of $\Delta \beta_{r,i}=0.02$, with dotted lines denoting the edges of the synchronized (marked as I) and moderately unsynchronized regions (II). Frequency-pulling occurs in the regions above and below these lines (region II), as shown by the inset for the blue marker. However, as we will see below, it is still possible to find optimal solutions even without frequency locking.

We numerically integrated Eq.~(\ref{eq:delay}) 200 times for each $(\beta_r,\beta_i)$ tuple in the area of interest for the four coupling matrices in Fig.~\ref{fig:matrices}. We then used the phases of the steady-state solutions to calculate the global minimum probabilities (GMP) in each case. The GMPs are shown in Fig.~\ref{fig:result}, and firstly, we are focused on analyzing a regime with robust SHIL, \emph{i.e.}, region I. As expected, the M{\"o}bius ladder, being a simple and highly symmetrical graph, has the highest success probability, but we observe common features between all four that can aid in the search for ideal nonlinearities. From the first inspection, we observe a sharp transition across line $\beta_i=\nu_{th} \beta_r +const$, above which the GMP drops to zero in all cases. This region is marked as I.a in Fig.~\ref{fig:result}. Although all spins are synchronized with binarizing signal $\omega_e$, the whole system cannot settle into the global minimum.

Below the transition line, in the region marked as I.b, the success probability never drops to zero with a clear trend to increase as $\beta_i$ decreases. This behavior is common for all tested problems, as shown in figs.~\ref{fig:result}b, c, and d. Hence, while it can be easier in practice to increase an amplifier gain compression $\beta_r$ to cross from region I.a to I.b, the proper tuning of the frequency nonlinearity $\beta_i$ is more beneficial and leads to a higher GMP. This increase of GMP with decreasing $\beta_i$ can be understood in terms of the locking bandwidth of the synchronization with respect to $\omega_e$. In general, the desynchronization of an oscillator with respect to a periodic driving signal increases the force that couples the two. In our model, the locking bandwidth with respect to $\omega_e$ increases with $\beta_i$, thus increasing this synchronizing force. On the other hand, given that all oscillators have approximately the same natural frequency, $\beta_i$ has no effect over their mutual force. For a large enough $\beta_i$, the force associated with the first term in Eq.~\ref{eq:coupling-sum} is larger than that of the second, and the individual oscillators are driven solely by the binarizing signal, thus flattening their energy landscape and hindering GMP.

More interestingly, for the same graphs, there is an increase in GMP in the vicinity of the edge of the synchronized region I.b for $\beta_i<0$, marked by the lower dashed line. This result is somewhat surprising, as one would intuitively aim for synchronization with $\omega_e$ to ensure binarization. However, as stated in the last paragraph, overcoupling to the $\omega_e$ leads to a flattening of the landscape. On the edge of synchronization, the spins can flip their sign the easiest as the force felt by the oscillators from the binarizing signal and the sum of the pairwise couplings (the two terms in Eq.~\ref{eq:coupling-sum}) is approximately the same. This ease of movement through the configuration space allows the machine to be attracted and locked to the global minimum more often, thus improving the GMP. Similar to the case for the $\nu_{th}$, one can tune the machine for successful operation with a variable gain amplifier~\cite{litvinenko2023SAWIM}.

In all cases, $|\beta_i|<0.2$ and the variable $\beta_r$ assure convergence to the optimal minimum with GMP$>0$. A detailed example of the differences in the evolution of the spin phases for different values of nonlinearities can be seen in Fig.~S1 of the supplementary materials. In the entire synchronized region I, this evolution is initially quite swift towards the stationary state. However, the final state does not correspond to the global minimum at the region I.a. For completeness, we also introduced a small difference in the natural frequencies of the oscillators to emulate the imprecision of physical devices. We found that a small variation in the individual oscillator frequencies does not meaningfully impact the operation or GMP of the model. These results are shown in more detail in the supplemental material Fig.~S2.

An interesting phenomenon occurs at low $\beta_r$ and high $\beta_i$ in region II, where the regime of frequency pulling instead of synchronization is realized by the binarizing signal, see Fig.~\ref{fig:coupling-search}. However, we found certain areas with non-zero, and even very high, GMP. Observing the \textit{relative} phases between spins, we found that they settle in the ground state locked to each other but change with respect to $\omega_e t$. This suggests that the coupling between spins is sufficiently robust to generate and maintain binarization even in the presence of frequency pulling. However, this alternative solution region seems to be very problem dependent; as we see that while the red and orange region (GMP$>$60\%) is quite large for the M{\"o}bius ladder in \ref{fig:result}(a), it becomes a smaller and less successful white region (GMP$<$15\%) for the non-symmetric graphs of Fig.~\ref{fig:result}(b-d), which have more complicated energy landscapes.

\begin{figure}[hbt!]
\centering
	\includegraphics[width=\linewidth]{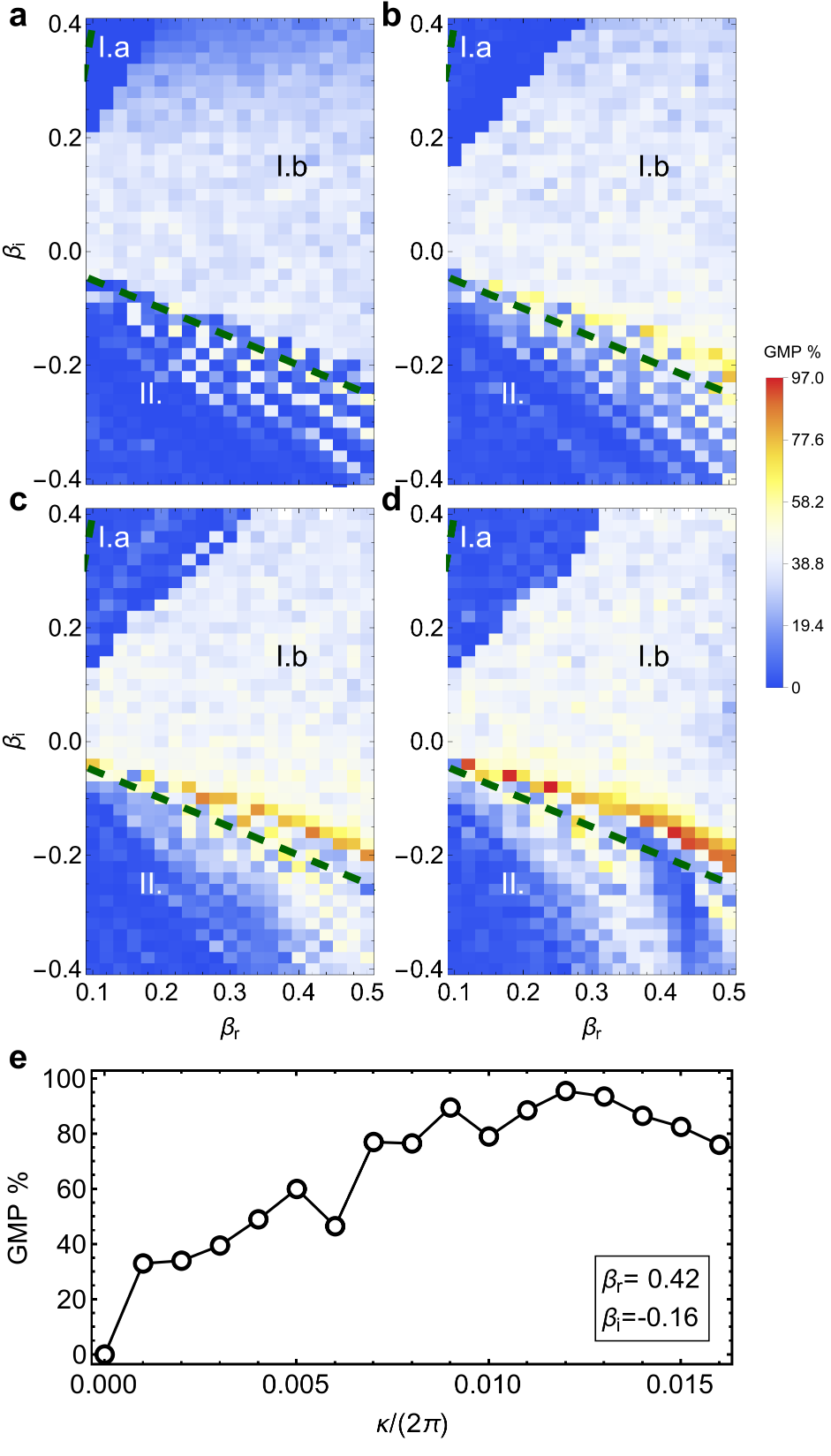}
	\caption{\textbf{GMP for different values of $\kappa$ for graph d.} $\kappa / 2 \pi = \text{(a)} \ 0.003, \ \text{(b)} \ 0.006, \ \text{(c)} \ 0.009, \ \text{(d)} \ 0.012$. The average GMP increases with $\kappa$ and can be optimally tuned for negative $\beta_i$. \ \text{(e)} \ GMP as a function of the coupling strength $\kappa$ for an optimal set $(\beta_r, \beta_i)$. There is a certain optimal value of $\kappa$ for which the probability of finding the global minimum is maximized.}
	\label{fig:dispersion}
\end{figure}

We also probed into the effect of the feedback coupling strength $\kappa$ over the GMP. It has been shown analytically~\cite{Khairul2023-stability-of-OIMs} and experimentally~\cite{litvinenko2023SAWIM} that this strength has an impact on the convergence of a system of coupled oscillators to a stable solution and the overall quality of the found solutions. We indeed found that the GMP increases substantially with $\kappa$ up to a certain optimal value throughout the whole region I.b, as seen from Fig.~\ref{fig:dispersion}, whereas the region I.a slightly expands, and I.b shrinks with $\kappa$. More remarkably, the GMP enhancement by optimal  $(\beta_r,\beta_i)$ at the lower edge of the region I.b becomes more prominent at higher $\kappa$, with GMR reaching $97\%$. From Fig.~\ref{fig:dispersion}.e, where GMP for the preeminent set $(\beta_r=0.42,\beta_i=-0.16)$ are shown, we can estimate $\kappa\simeq 0.012$ as the optimal value that gives the highest GMP. Therefore, one can conclude that \emph{there is an optimal set of  $(\kappa,\beta_r,\beta_i)$ for the best performance of time-multiplexed IM}, wherein $(\beta_r,\beta_i)$ lays close to the edge of the synchronization region. 
Thus, for negative $\beta_i$, one can take advantage of the behavior at the synchronization edge to tune the $\beta_r$ and $\kappa$ to maximize the GMP. The impact of coupling strength $\kappa$ has been studied previously with similar results~\cite{Khairul2023-stability-of-OIMs} where the authors showed that there is a certain optimal value, for which all global minima are stable state solutions of the oscillator network. These features can improve the design of hardware-based computation acceleration, such as in time-multiplexed IM, where $\kappa$ can be easily changed with the introduction of a variable attenuating element~\cite{litvinenko2023SAWIM} to improve the GMP dramatically.

\begin{figure}[hbt!]
\centering
	\includegraphics[width=\linewidth]{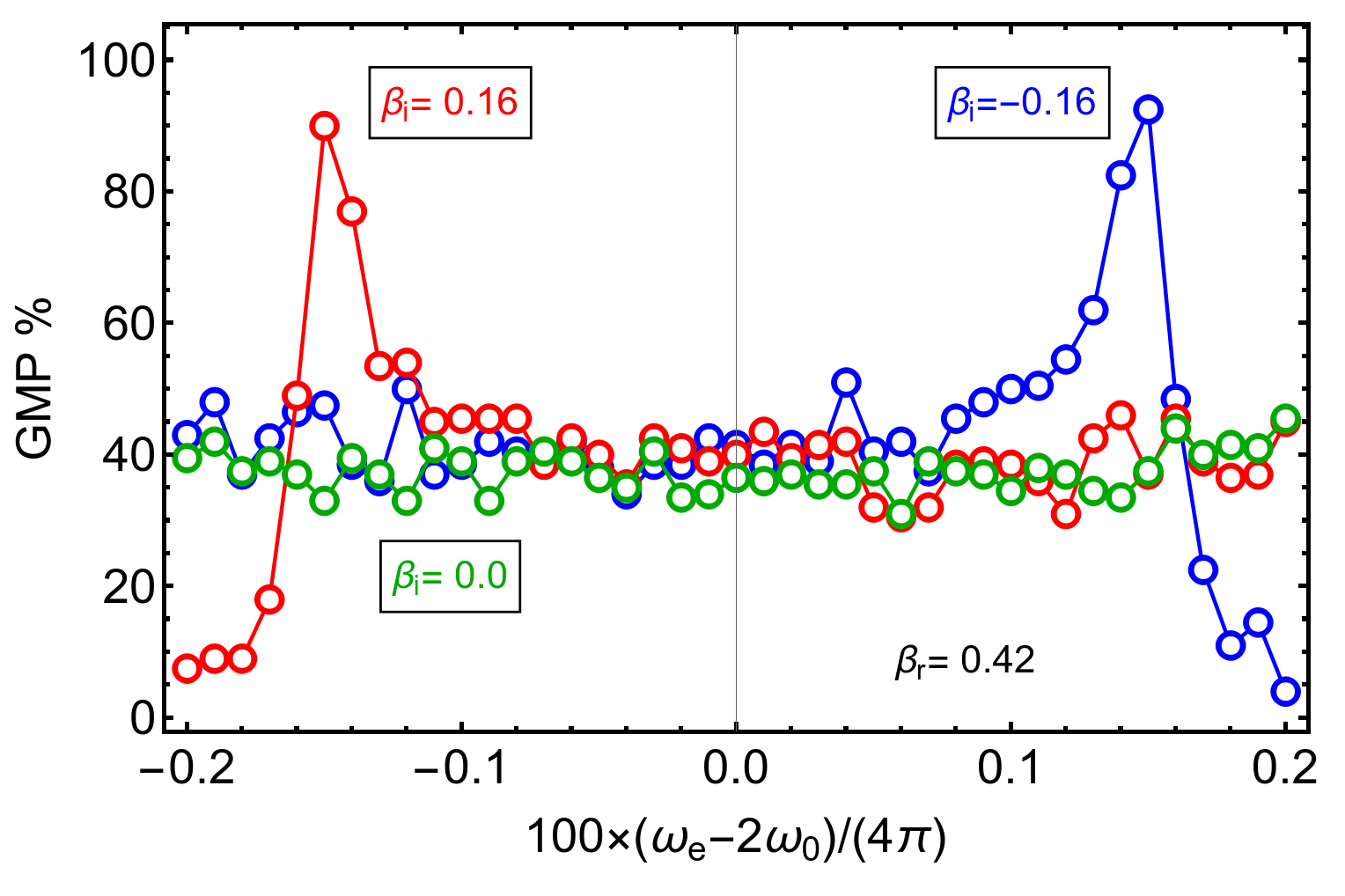}
	\caption{\textbf{GMP dependence on $\omega_e$ for graph d.} Since changing the sign of $\beta_i$ is not always possible, adjusting the detuning between the binarizing signal $\omega_e$ and the natural frequency of the oscillators $\omega_0$ is yet another control mechanism for optimizing the GMP.}
	\label{fig:ext_freq}
\end{figure}

Finally, since one may not be able to choose the sign for $\beta_i$ for a given machine, we explored the effect of the detuning between $\omega_e$ and $\omega_0$ on GMP. In Fig.~\ref{fig:ext_freq}, we observe that GMP enhancement by optimal $\beta_i$ is symmetrical with respect to this detuning and can be optimized by changing the sign of $\omega_e$. However, to achieve a high GMP, it is necessary to use an optimal value of the frequency nonlinearity. Thus, no GMP enhancement is observed for $\beta_i=0$ (see green line in Fig.~\ref{fig:ext_freq}). This result is consistent with the operational principle of time-multiplexed IMs, as the nonlinearity allows for the exploration of the configuration space and eventual convergence to the global minimum. 

\section{Conclusions}

Our model of driven coupled oscillators with a delay loop, described by Eqs.~(\ref{eq:delay}) and (\ref{eq:coupling-sum}), works well to predict the dynamics of time-multiplexed IMs. We used numerical integration to investigate the effect of the compression gain $\beta_r$ and nonlinear frequency shift $\beta_i$ over the solution space of four different coupling topologies with different symmetries, densities, and optimal solutions. In all cases, we found a sharp transition across the line $\beta_i=\nu_{th} \beta_r +const$, with $\nu_{th}$ and the constant being characteristic of each topology. The optimal solution region lies below this line.

We also found that it is possible to leverage the system's dynamics at the edge of synchronization with either positive or negative frequency nonlinearity $\beta_i$ by tuning the compression gain (amplitude nonlinearity) $\beta_r$ and by controlling the detuning of the binarizing signal $\omega_e$ and the oscillators' natural frequency $\omega_0$. A combination of both methods makes for plausible optimization methods for GMP in physical time-multiplexed IMs. These results largely remain unchanged when we introduce a small amount of variance between the oscillators' natural frequencies, showing the approach's robustness and applicability for more realistic case studies.

We studied limiting cases for the two terms in our driving force: the binarizing signal $K_e$ and the coupling strength between oscillators $\kappa$. For the former, we found that the network can find optimal solutions even if the oscillators are subject to a small amount of frequency pulling and are not fully synchronized with the binarizing signal $\omega_e$. For the latter, we showed that $\kappa$ can increase the average success probability up to a certain optimal strength, albeit with a slight reduction of the convergence region. These results align with previous analytical and experimental work and can be used to formulate optimal operation conditions for a wide range of time-multiplexed IMs. As interest grows in this field, finding the correct operational regimes for obtaining results comparable to conventional computing architectures with minimum overhead is becoming ever more important for developing this evolving computation accelerator.

\section{Acknowledgements}
This work was supported by a Knut and Alice Wallenberg Foundation WALP grant, KAW 253129326, a Horizon 2020 research and innovation programme ERC Advanced Grant No.~835068 ``TOPSPIN'', an ERC Proof of Concept Grant No.~101069424 ``SPINTOP'', and the Marie Skłodowska-Curie grant agreement No.~101111429 ``SWIM''.

\nocite{*}

\bibliography{main}% Produces the bibliography via BibTeX.

\end{document}

\preprint{APS/123-QED}

\title{Supplementary material for \\A numerical model for time-multiplexed Ising machines\\based on delay-line oscillators}

\author{Roman V. Ovcharov}
\affiliation{ 
Department of Physics, University of Gothenburg, Gothenburg 41296, Sweden
}

\author{Victor H. González}
\affiliation{ 
Department of Physics, University of Gothenburg, Gothenburg 41296, Sweden
}

\author{Artem Litvinenko}%
\affiliation{ 
Department of Physics, University of Gothenburg, Gothenburg 41296, Sweden
}

\author{Johan \AA kerman}
\affiliation{ 
Department of Physics, University of Gothenburg, Gothenburg 41296, Sweden
}
\affiliation{
Center for Science and Innovation in Spintronics and Research Institute of Electrical Communication, Tohoku University, Aoba-ku, Sendai 980-8577, Japan
}

\author{Roman S.  Khymyn}%
\affiliation{ 
Department of Physics, University of Gothenburg, Gothenburg 41296, Sweden
}

\date{\today}
\maketitle

\begin{figure}[hbt!]
\centering
	\includegraphics[width=\linewidth]{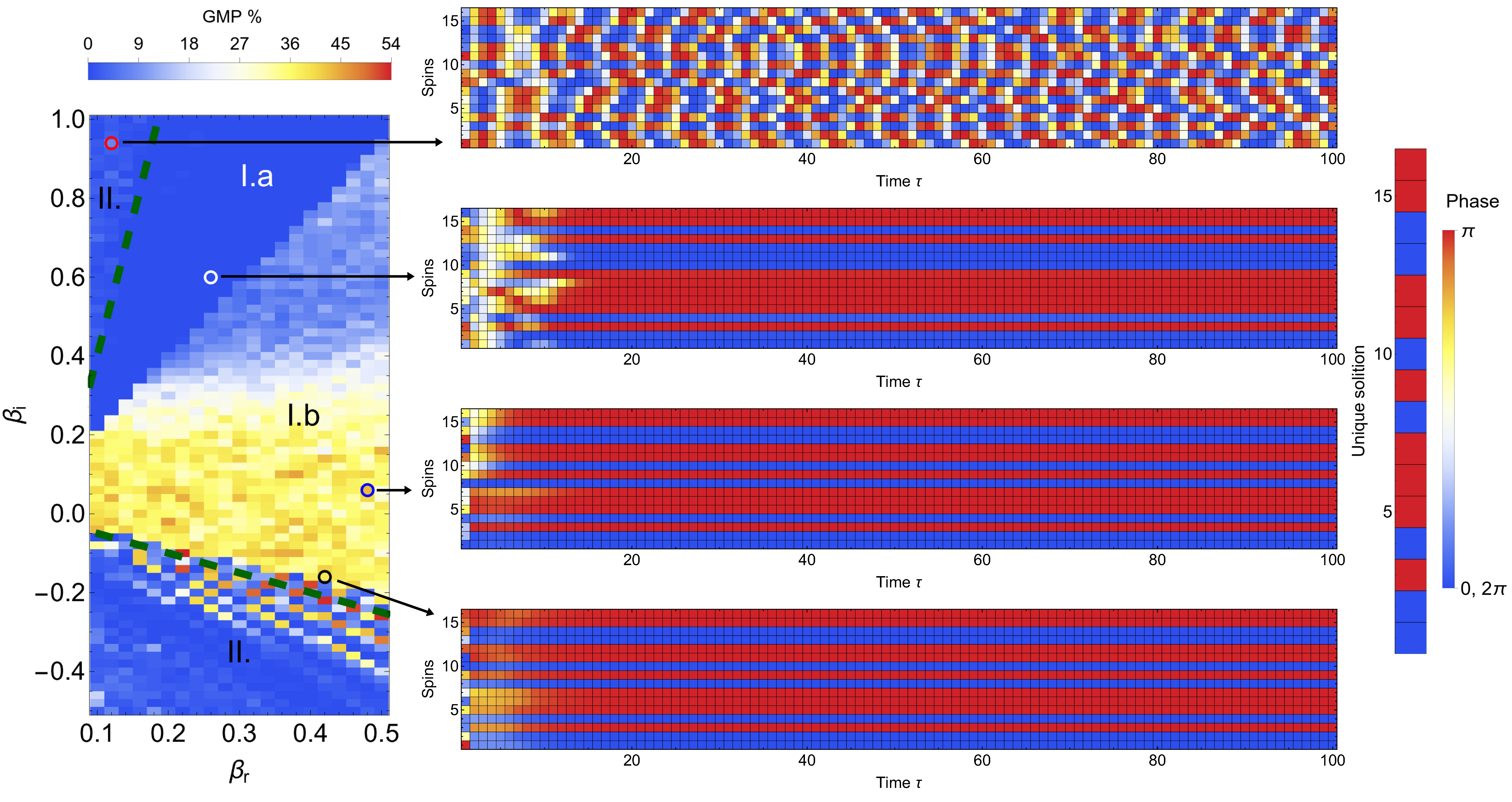}
	\caption{\textbf{Time traces illustrating the phase dynamics of a 16-spin system.} The chosen graph corresponds to graph (d) in Fig. 1 of the main text. On the left, the global minimum probability is plotted as a function of $\beta_r$ and $\beta_i$, mirroring subfigure (d) in Fig. 3 of the main text. For four specific pairs of ($\beta_r$, $\beta_i$), the phase evolution of  16 oscillators is shown in color. The two subfigures below showcase the optimal region ($\beta_r=0.42$, $\beta_i=-0.16$) and ($\beta_r=0.48$, $\beta_i=0.06$), where the phases rapidly converge to the correct configuration with the highest possible probability and remain stable. In the subfigure above ($\beta_r=0.26$, $\beta_i=0.6$), the solution remains unchanged over time but always gets stuck in local minima. The upper subfigure corresponds to the frequency-pulling region ($\beta_r=0.12$, $\beta_i=0.94$), where oscillator phases are not fixed; however, their relative phases can align with the global minimum solution. The rightmost subfigure shows the unique (excluding flipping all spins) solution corresponding to the global energy minimum.}
	\label{fig:phases_in_time}
\end{figure}

\begin{figure}[hbt!]
\centering
	\includegraphics[width=0.7\linewidth]{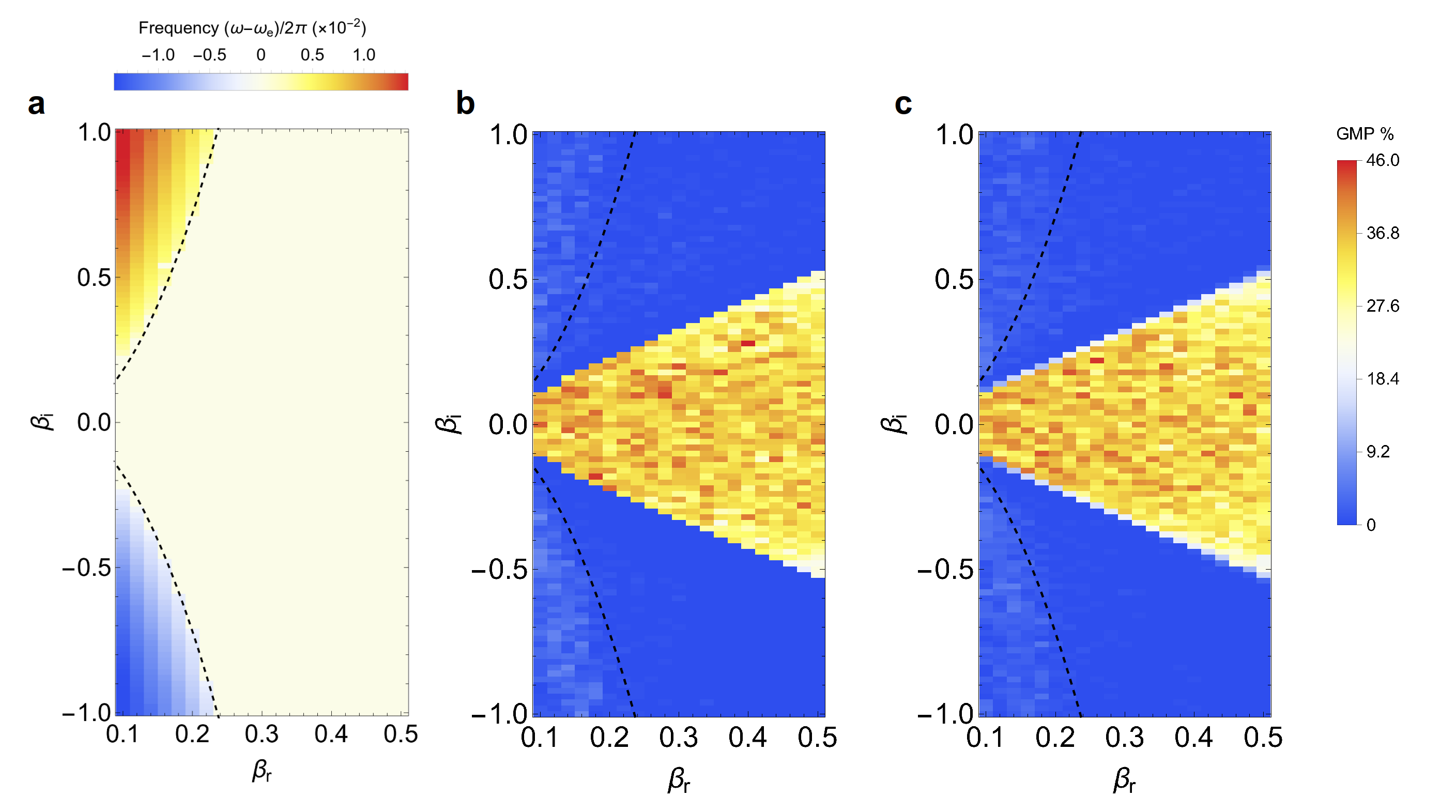}
	\caption{\textbf{Variability in individual oscillator frequencies.} The global minimum probabilities for graph d) in Fig. 1 of the main text under two scenarios: a) where all oscillators share the same natural frequency (copy of the subfigure (d) in Fig. 3 of the main text), and b) with a dispersion in values of $\omega_0$. In the latter scenario, each oscillator's frequency $\omega_0$ is set using Gaussian distribution with a mean of $\omega_0/2\pi = 1.0$ and a standard deviation of $\sigma=5.0\times10^{-4}$ before each iteration.}
	\label{fig:dispersion}
\end{figure}

\bibliography{main}